\RequirePackage{fixltx2e}
\documentclass[twoside,journal, letterpaper, final, twosided]{IEEEtran}
\usepackage[T1]{fontenc}
\usepackage[latin9]{inputenc}
\usepackage{textcomp}
\usepackage{graphicx}
\usepackage[unicode=true,pdfusetitle,
 bookmarks=true,bookmarksnumbered=false,bookmarksopen=false,
 breaklinks=false,pdfborder={0 0 1},backref=false,colorlinks=false]
 {hyperref}
\usepackage{breakurl}

\makeatletter
\setkeys{Gin}{width=\linewidth}
\usepackage{cite}

\long\def\@makecaption#1#2{\ifx\@captype\@IEEEtablestring%
\footnotesize\begin{center}{\normalfont\footnotesize #1}\\
{\normalfont\footnotesize\scshape #2}\end{center}%
\@IEEEtablecaptionsepspace
\else
\@IEEEfigurecaptionsepspace
\setbox\@tempboxa\hbox{\normalfont\footnotesize {#1.}~~ #2}%
\ifdim \wd\@tempboxa >\hsize%
\setbox\@tempboxa\hbox{\normalfont\footnotesize {#1.}~~ }%
\parbox[t]{\hsize}{\normalfont\footnotesize \noindent\unhbox\@tempboxa#2}%
\else
\hbox to\hsize{\normalfont\footnotesize\hfil\box\@tempboxa\hfil}\fi\fi}
\ifCLASSOPTIONcompsoc
  \usepackage[caption=false,font=normalsize,labelfont=sf,textfont=sf]{subfig}
\else
  \usepackage[caption=false,font=footnotesize]{subfig}
\fi

\usepackage{amsmath}

\usepackage{dblfloatfix}

\makeatother

\begin{document}

\title{\emph{In Vivo} Communications: Steps Toward the Next Generation of
Implantable Devices}

\author{\vspace{0mm}
Ali Fatih Demir, \emph{Student Member, IEEE}, Z. Esat Ankaral\i ,
\emph{Student Member, IEEE}, Qammer H. Abbasi, \emph{Member, IEEE},
Yang Liu, \emph{Student Member, IEEE}, Khalid Qaraqe, \emph{Senior
Member, IEEE}, Erchin Serpedin, \emph{Fellow, IEEE}, Huseyin Arslan,
\emph{Fellow, IEEE}, and Richard D. Gitlin, \emph{Life Fellow, IEEE}
\thanks{Manuscript received June 01, 2015, revised December 21, 2015,
and accepted January 10, 2016. This publication was made possible
by NPRP grant \# 6-415-3-111 from the Qatar National Research Fund.}
\thanks{A. F. Demir, Z. E. Ankaral\i , Y. Liu, H. Arslan, and R.
D. Gitlin are with the Department of Electrical Engineering, University
of South Florida, Tampa, FL 33620 USA (e-mail: afdemir@mail.usf.edu;
zekeriyya@mail.usf.edu; yangl@mail.usf.edu; arslan@.usf.edu; richgitlin@usf.edu).}\thanks{
Q. H. Abbasi and K. Qaraqe are with the Department of Electrical and
Computer Engineering, Texas A\&M University at Qatar, Doha 23874,
Qatar (email: qammer.abbasi@qatar.tamu.edu; khalid.qaraqe@qatar.tamu.edu).
}\thanks{E. Serpedin is with the Department of Electrical and Computer
Engineering, Texas A\&M University, College Station, TX 77843 USA
(email: serpedin@tamu.edu).} }
\maketitle
\begin{abstract}
\textit{In vivo} wireless medical devices have the potential to play
a vital role in future healthcare technologies by improving the quality
of human life. In order to fully exploit the capabilities of such
devices, it is necessary to characterize and model the \textit{in
vivo} wireless communication channel. Utilization of this model will
have a significant role in improving the communication performance
of embedded medical devices in terms of power, reliability and spectral
efficiency. In this paper, the state of the art in this field is presented
to provide a comprehensive understanding of current models. Such knowledge
will be used to optimize the design and selection of various \textit{in
vivo} wireless communication methods, operational frequencies, and
antenna design. Finally, open research areas are discussed for the
future studies. 
\end{abstract}


\markboth{IEEE Vehicular Technology Magazine}{Demir \MakeLowercase{\textit{et al.}}:In Vivo Communications: Steps Toward the Next Generation of Implantable Devices}

\section{Introduction}

Technological advances in biomedical engineering have significantly
improved the quality of life and increased the life expectancy of
many people. One component of such advanced technologies is represented
by wireless \textit{in vivo} sensors and actuators, e.g., pacemakers,
internal drug delivery devices, nerve stimulators, wireless capsule
endoscopes (WCEs), etc. \textit{In vivo}-wireless body area networks
(WBANs) \cite{IEEE_wban} and their associated technologies are the
next step in this evolution and offer a cost efficient and scalable
solution along with the integration of wearable devices. \textit{In
vivo} WBAN devices are capable of providing continuous health monitoring
and reducing the invasiveness of surgery. Furthermore, the patient
information can be collected over a larger period of time and physicians
are able to perform more reliable analysis by exploiting this \textit{big
data} rather than relying on the data recorded in short hospital visits
\cite{Kiourti,Movassaghi}.

To fully exploit and increase further the potential of WBANs for practical
applications, it is necessary to accurately assess the propagation
of electromagnetic (EM) waveforms in an \textit{in vivo} communication
environment (implant-to-implant and implant-to-external device) and
obtain accurate channel models that are necessary to optimize the
system parameters and build reliable, efficient, and high-performance
communication systems. In particular, creating and assessing such
a model is necessary for achieving high data rates, target link budgets,
determining optimal operating frequencies, and designing efficient
antennas and transceivers including digital baseband transmitter/receiver
algorithms. Therefore, investigation of the \textit{in vivo} wireless
communication channel is crucial to obtain a better performance for
\textit{in vivo}-WBAN devices and systems. Although, on-body wireless
communication channel characteristics have been well investigated
\cite{Movassaghi}, there are relatively few studies of \textit{in
vivo} wireless communication channels.

While there exist multiple approaches to \textit{in vivo} communications,
in this paper we will focus on EM communications. Since the EM wave
propagates through a very lossy environment inside the body and predominant
scatterers are present in the near-field region of the antenna, \textit{in
vivo} channel exhibits different characteristics than those of the
more familiar wireless cellular and Wi-Fi environments. In this paper,
we present the state-of-the-art of \textit{in vivo} channel characterization
and discuss several research challenges by considering various communication
methods, operational frequencies, and antenna designs.

The rest of the paper is organized as follows. In Section II, EM modeling
of the human body is reviewed, which is essential for \textit{in vivo}
wireless communication channel characterization. Section III discusses
EM wave propagation through human tissues. Section IV presents the
choice of operational frequencies based on current standards and discusses
their effects on the communication system performance. In Section
V, the challenges of \textit{in vivo} antenna design are briefly discussed
as the antenna is generally considered to be an integral part of the
\textit{in vivo} channel. Section VI reviews the propagation models
for the \textit{in vivo} wireless communication channel and discusses
the main differences relative to the \textit{ex vivo} channel. In
Section VII, several open research problems and future research directions
are addressed. The last section summarizes our observations and conclusions.

\section{EM Modeling of the Human Body}

In order to investigate the \textit{in vivo} wireless communication
channel, accurate body models and knowledge of the electromagnetic
properties of the tissues are crucial. Human autopsy materials and
animal tissues have been measured over the frequency range from 10
Hz to 20 GHz \cite{Hall} and the frequency-dependent dielectric properties
of the tissues are modeled by four-pole Cole-Cole equation, which
is expressed as:

\begin{equation}
\epsilon(\omega)=\epsilon_{\infty}+\sum_{m=1}^{4}\frac{\Delta\epsilon_{m}}{1+(j\omega\tau_{m})^{(1-\alpha_{m})}}+\frac{\sigma}{j\omega\epsilon_{0}}\label{eq:Cole-Cole}
\end{equation}
where $\epsilon_{\infty}$ stands for the body material permittivity
at terahertz frequency, $\epsilon_{0}$ denotes the free-space permittivity,
$\sigma$ represents the ionic conductivity and $\epsilon_{m}$, $\tau_{m}$,
$\alpha_{m}$ are the body material parameters for each anatomical
region. The electromagnetic properties such as conductivity, relative
permittivity, loss tangent, and penetration depth can be derived using
these parameters in Eq. \ref{eq:Cole-Cole}.

\begin{figure}[!b]
\centering\includegraphics[width=3.2in]{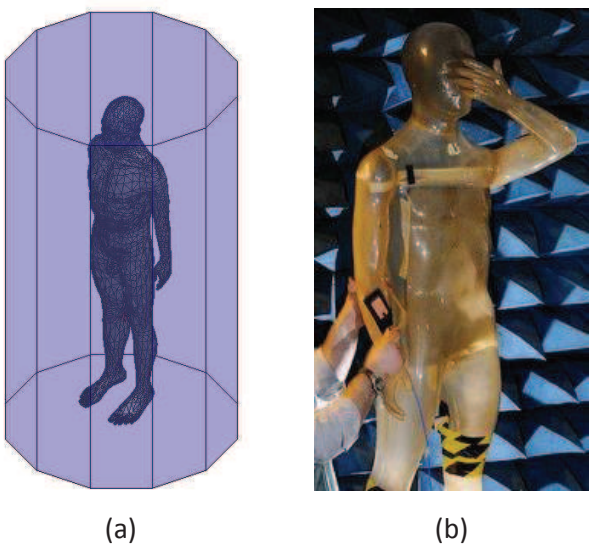} \caption{Heterogeneous human body models: (a) HFSS\protect\textsuperscript{\textregistered{}}
model, (b) physical phantom \cite{Alomainy}.\label{fig:system}}
\end{figure}

Various physical and numerical phantoms have been designed in order
to simulate the dielectric properties of the tissues for experimental
and numerical investigation. These can be classified as homogeneous,
multi-layered and heterogeneous phantom models. Although, heterogeneous
models provide more realistic approximation to the human body, design
of physical heterogeneous phantoms is quite difficult and performing
numerical experiments on these models is very complex and resource
intensive. On the other hand, homogeneous or multi-layer models cannot
differentiate the EM wave radiation characteristics for different
anatomical regions. Fig. \ref{fig:system} shows examples of heterogeneous
physical and numerical phantoms.

Analytical methods are generally viewed as infeasible and require
extreme simplifications. Therefore, numerical methods are used for
characterizing the \textit{in vivo} wireless communication channel.
Numerical methods provide less complex and appropriate approximations
to Maxwell's equations via various techniques, such as uniform theory
of diffraction (UTD), finite integration technique (FIT), method of
moments (MoM), finite element method (FEM) and finite-difference time-domain
method (FDTD). Each method has its own pros and cons and should be
selected based on the simulation model and size, operational frequency,
available computational resources and interested characteristics such
as power delay profile (PDP), specific absorption rate (SAR), etc.
A detailed comparison for these methods is available in \cite{Hall}
and \cite{pellegrini2013antennas}.

It may be preferable that numerical experiments should be confirmed
with real measurements. However, performing experiments on a living
human is carefully regulated. Therefore, anesthetized animals \cite{Lee,chavez2013experimental}
or physical phantoms, allowing repeatability of measurement results,
\cite{Alomainy,Lin} are often used for experimental investigation.
In addition, the first study conducted on a human cadaver was reported
in \cite{demir_vtc}.

\section{EM Wave Propagation Through Human Tissues}

Propagation in a lossy medium, such as human tissues, results in a
high absorption of EM energy. The absorption effect varies with the
frequency dependent electrical characteristics of the tissues, which
mostly consist of water and ionic content \cite{Yazdandoost}. The
specific absorption rate (SAR) provides a metric for the absorbed
power amount in the tissue and is expressed as follows:

\begin{equation}
SAR=\frac{\sigma|E|^{2}}{\rho}\label{Eq:SAR}
\end{equation}
where $\sigma$, ${\rm E}$ and $\rho$ represent the conductivity
of the material, the RMS magnitude of the electric field and the mass
density of the material, respectively. The Federal Communications
Commission (FCC) recommends the SAR to be less than 1.6 W/kg taken
over a volume having 1 gram of tissue \cite{Ketterl}.

\begin{figure}
\centering\includegraphics[width=3.2in]{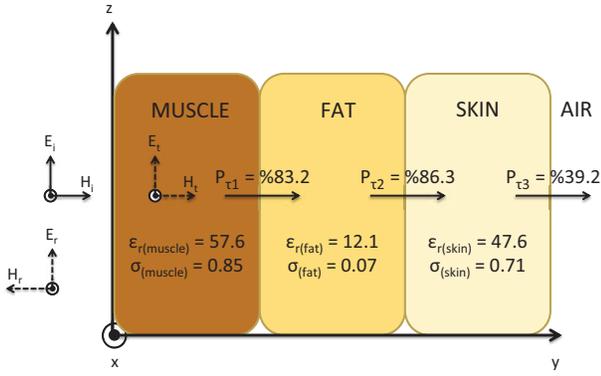} \caption{Multi-layer human tissue model at 403 MHz ($\epsilon_{r}$: Permittivity,
$\sigma$: Conductivity, $P_{\tau}$: Power transmission factor).\label{fig:multiLayer}}
\end{figure}

When an EM plane wave propagates through the interface of two media
having different electrical properties, its energy is partially reflected
and the remaining portion is transmitted through the boundary of these
mediums. Superposition of the incident and the reflected wave can
cause a standing wave effect that may increase the SAR values \cite{Yazdandoost}.
A multi-layer tissue model at 403 MHz, where each layer extends to
infinity (much larger than the wavelength of EM waves) is illustrated
in Fig. \ref{fig:multiLayer}. The dielectric values and power transmission
factors of this model were calculated in \cite{scanlon2003analysis}.
If there is a high contrast in the dielectric values of mediums/tissues,
wave reflection at the boundary increases and transmitted power decreases.
The limitations on communications performance imposed by the SAR limit
have been investigated in \cite{Ketterl}.

In addition to the absorption and reflection losses, EM waves also
suffers from expansion of the wave fronts (which assume an ever-increasing
sphere shape from an isotropic source in free space), diffraction
and scattering (which depend on the EM wavelength). Section VI discusses
\textit{in vivo} propagation models by considering these effects in
detail.

\begin{table}[b]
\caption{Frequency bands and bandwidths for the three different propagation
methods in IEEE 802.15.6 \label{fig:freqTable}}

\centering\includegraphics[width=1\columnwidth]{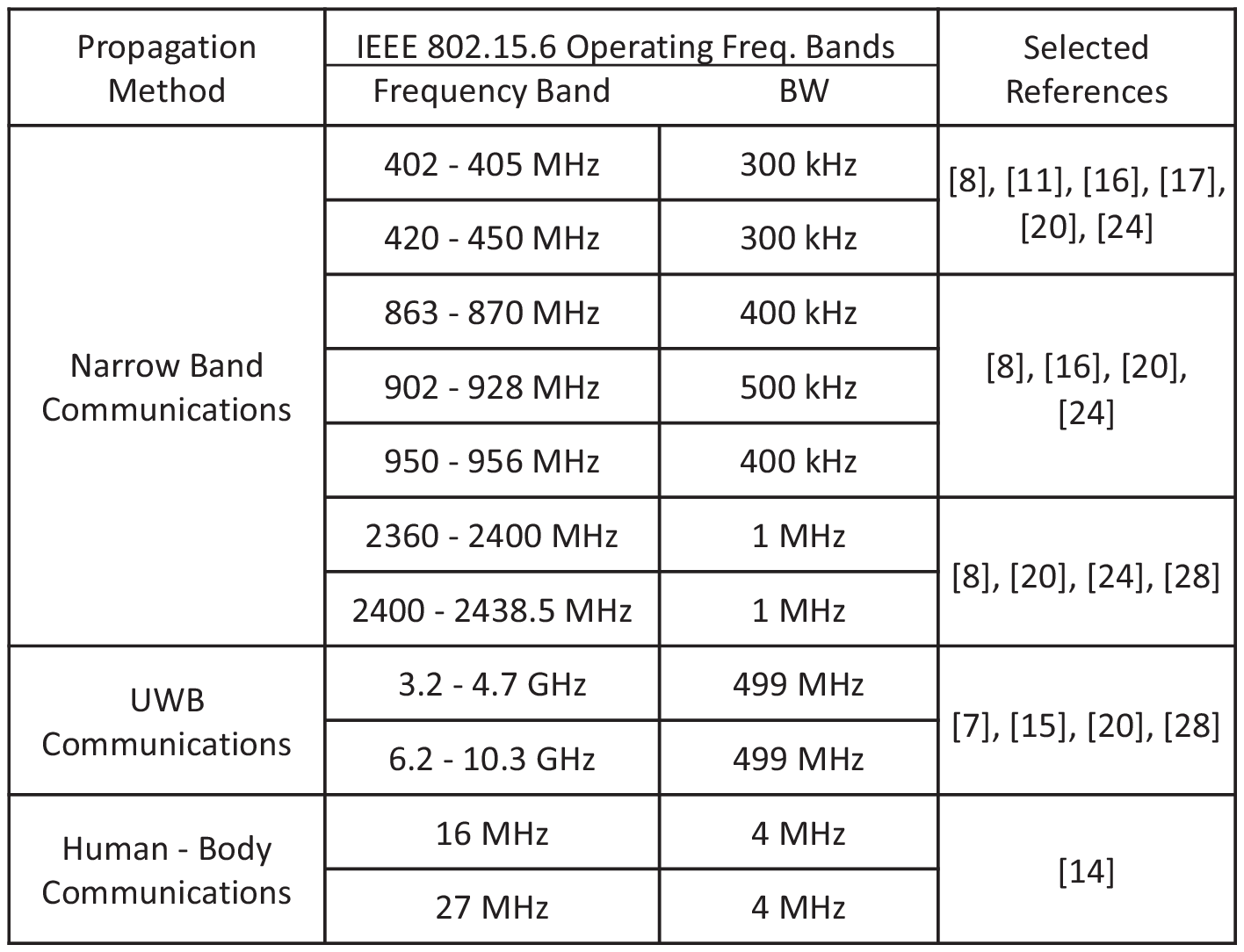}
\end{table}

\section{Frequency of Operation}

Since EM waves propagate through the frequency dependent materials
inside the body, the operating frequency has an important effect on
the communication channel. Accordingly, we summarize the allocated
and recommended frequencies including their effects for \textit{the
in vivo} wireless communications in this section.

The IEEE 802.15.6 standard \cite{IEEE_wban} was released in 2012
to regulate short-range wireless communications inside or in the vicinity
of the human body, and are classified as human-body communications
(HBC) \cite{wegmueller}, narrow band (NB) communications, and ultra-wide
band communications (UWB). The frequency bands and channel bandwidths
(BW) allocated for these communication methods are summarized in Table
I. An IEEE 802.15.6 compliant \textit{in vivo}-WBAN device must operate
in at least one of these frequency bands.

NB communications operates at lower frequencies compared to UWB communications
and hence suffer less from absorption. This can be appreciated by
considering Eq. \ref{eq:Cole-Cole} and Eq. \ref{fig:multiLayer}
that describe the absorption as a function of frequency. The medical
device radio communications service (MedRadio uses discrete bands
within the 401-457 MHz spectrum including international medical implant
communication service (MICS) band) and medical body area network (MBAN,
2360-2400 MHz) are allocated by the FCC for medical devices usage.
Therefore, co-user interference problems are less severe in these
frequency bands. However, NB communications are only allocated small
bandwidths (1 MHz at most) in the standard as shown in Table I. The
IEEE 802.15.6 standard does not define a maximum transmit power and
the local regulatory bodies limit it. The maximum power is restricted
to 25 $\mu$W EIRP (Equivalent Isotropic Radiated Power) by FCC, whereas
it is set to 25 $\mu$W ERP (Equivalent Radiated Power) by ETSI (European
Telecommunication Standards Institute) for the 402-405 MHz band.

UWB communications is a promising technology to deploy inside the
body due to its inherent features including high data rate capability,
low power, improved penetration (propagation) abilities through tissues
and low probability of intercept. The large bandwidths for UWB (499
MHz) enable high data rate communications and applications. Also,
UWB signals are inherently robust against detection and smart jamming
attacks because of their extremely low maximum EIRP spectral density,
which is -41.3 dBm/Mhz \cite{chavez2013propagation}. On the other
hand, UWB communications inside the body suffer from pulse distortion
caused by frequency dependent tissue absorption and the limitations
imposed by compact antenna design.

\section{\textit{In Vivo} Antenna Design Considerations}

Unlike free-space communications, \textit{in vivo} antennas are often
considered to be an integral part of the channel and they generally
require different specifications than the \textit{ex vivo} antennas
\cite{Hall,Sani,Sayrafian,Bahrami}. In this section, we will describe
their salient differences as compared to \textit{ex vivo} antennas.

\textit{In vivo} antennas are subject to strict size constraints and
in addition, they need to be bio-compatible. Although, copper antennas
have better performance, only specific types of materials such as
titanium or platinum should be used for \textit{in vivo} communications
due to their noncorrosive chemistry \cite{Movassaghi}. The standard
definition of the gain is not valid for \emph{in vivo} antennas since
it includes body effects \cite{kim2004implanted}. As noted above,
the gain of the \emph{in vivo} antennas cannot be separated from the
body effects as the antennas are considered to be an integral part
of the channel. Hence, the \textit{in vivo} antennas should be designed
and placed carefully. When the antennas are placed inside the human
body, their electrical dimensions and gains decrease due to the high
dielectric permittivity and high conductivity of the tissues, respectively
\cite{Gemio}. For instance, fat has a lower conductivity than skin
and muscle. Therefore, \textit{in vivo} antennas are usually placed
in a fat (usually subcutaneous fat) layer to increase the antenna
gain. This placement also provides less absorption losses due to shorter
propagation path. However, the antenna size becomes larger in this
case. In order to reduce high losses inside the tissues, a high permittivity,
low loss coating layer can be used. As the coating thickness increases,
the antenna becomes less sensitive to the surrounding material \cite{Gemio}. 

\begin{figure}
\centering\includegraphics[width=3.5in]{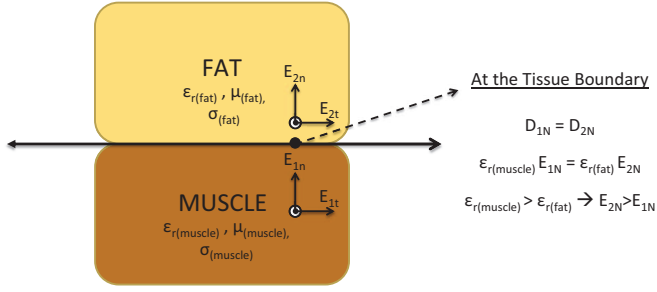} \caption{EM propagation through tissue interface. \label{aa}}
\end{figure}

Lossy materials covering the \textit{in vivo} antenna change the electrical
current distribution in the antenna and radiation pattern. It is reported
in \cite{Sani} that directivity of \emph{in vivo} antennas increases
due to curvature of body surface, losses and dielectric loading from
the human body. Therefore, this increased directivity should be taken
into account as well in order not to harm the tissues in the vicinity
of the antenna.

\begin{figure}[!b]
\centering\includegraphics[width=1\columnwidth]{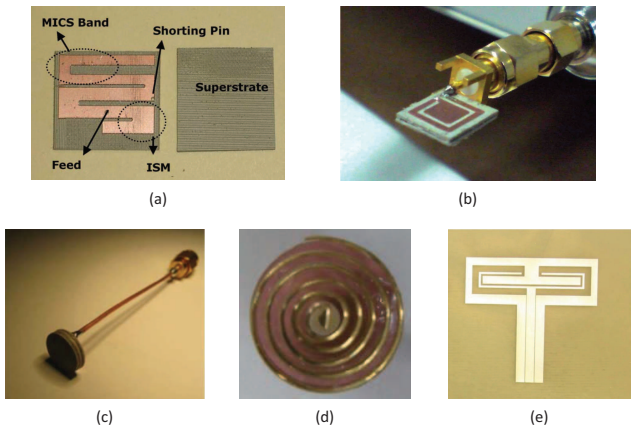} \caption{Selected \emph{in vivo} antenna samples: a) A dual-band implantable
antenna \cite{a1}, b) A miniaturized implantable broadband stacked
planar inverted-F antenna (PIFA) \cite{a2}, c) A miniature scalp-implantable
\cite{Kiourti}, d) A wideband spiral antenna for WCE \cite{Lee},
e) An implantable folded slot dipole antenna \cite{a5}. \label{az}}
 
\end{figure}

\textit{In vivo} antennas can be classified into two main groups as
electrical and magnetic antennas. Electrical antennas, e.g., dipole
antennas, generate electric fields (E-field) normal to the tissues,
while magnetic antennas, e.g., loop antennas produce E-fields tangential
to the human tissues \cite{Yazdandoost}. Normal E-field components
at the medium interfaces overheat the tissues due to the boundary
condition requirements as illustrated in Fig. \ref{aa}. The muscle
layer has a larger permittivity value than the fat layer and hence,
the E-field increases in the fat layer. Therefore, magnetic antennas
allow higher transmission power for \textit{in vivo}-WBAN devices
as can be understood from Eq. 2. In practice, magnetic loop antennas
require large sizes, which is a challenge to fit inside the body.
Accordingly, smaller size spiral antennas having a similar current
distribution as loop antennas can be used for \textit{in vivo} devices
\cite{Lee}. Representative antennas designed for \textit{in vivo}
communications are shown in Fig. \ref{az}.

\section{\textit{In Vivo} EM Wave Propagation Models}

Up to this point, important factors for \textit{in vivo} wireless
communication channel characterization, such as EM modeling of the
human body, propagation through the tissues, , selection of the operational
frequencies, and \emph{in vivo} antenna design considerations have
been reviewed. In this section, we will focus on EM wave propagation
inside the human body considering the anatomical features of organs
and tissues. Then, the analytical and statistical path loss models
will be presented. Since the EM wave propagates through a very lossy
environment inside the body and predominant scatterers are present
in the near-field region of the antenna, \textit{in vivo} channel
exhibits different characteristics than those of the more familiar
wireless cellular and Wi-Fi environments.

\begin{figure}[!t]
\centering \includegraphics[width=0.9\columnwidth]{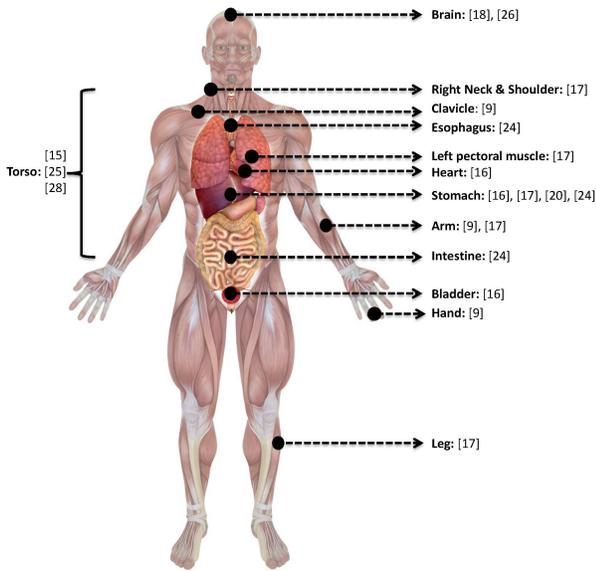}
\caption{Investigated anatomical human body regions. \label{fig:aa4}}
\end{figure}

EM wave propagation inside the body is subject-specific and strongly
related to the location of antenna as demonstrated in \cite{Basar,Lin,Sani}
and \cite{demir}. Therefore, channel characterization is mostly investigated
for a specific part of the human body. Fig. \ref{fig:aa4} shows several
investigated anatomical regions for various \textit{in vivo}-WBAN
applications and Table II provides further details about these studies.
For example, the heart area has been studied for implantable cardioverter
defibrillator and pacemakers, while the gastrointestinal tract (GI)
including esophagus, stomach and intestine has been investigated for
WCE applications. The bladder region is studied for wirelessly controlled
valves in the urinary tract and the brain is investigated for neural
implants \cite{chenz,Bahrami}. Also, clavicle, arm and hands are
specifically studied as they are affected less by the \textit{in vivo}
medium.

When the \textit{in vivo} antenna is placed in an anatomically complex
region, path loss, a measure of average signal power attenuation,
increases \cite{Basar}. This is the case with the intestine which
presents a complex structure with repetitive, curvy-shaped, dissimilar
tissue layers, while the stomach has a smoother structure. As a result,
the path loss is greater in the intestine than in the stomach even
at equal \textit{in vivo} antenna depths.

Various analytical and statistical path loss formulas have been proposed
for the \textit{in vivo} channel in the literature as listed in Table
III. These formulas have been derived considering different shadowing
phenomena for the \textit{in vivo} medium. The initial three models
are functions of the Friis transmission equation \cite{Hall}, return
loss and absorption in the tissues. These models are valid, when either
the far field conditions are fulfilled or few scattering objects exist
between the transmitter and receiver antennas.

The free space path loss (FSPL) is expressed by the Friis transmission
equation in the first model in Table \ref{tab:PL_models}. The FSPL
mainly depends on gain of antennas, distance, and operating frequency.
Its dependency on distance is a result of expansion of the wave fronts
as explained in Section III. Additionally, FSPL is frequency dependent
due to the relationship between the effective area of the receiver
antenna and wavelength. The two equations of the FSPL model in Table
\ref{tab:PL_models} are derived including the antenna return loss
and absorption in the tissues. Another analytical model, PMBA \cite{Gupta},
calculates the SAR over the entire human body for the far and near
fields, and gives the received power using the calculated absorption.
Although, these analytical expressions provide intuition about each
component of the propagation models, they are not practical for link
budget design as similar to the wireless cellular communication environment.

\begin{figure}[!b]
 \centering\includegraphics[width=3.5in]{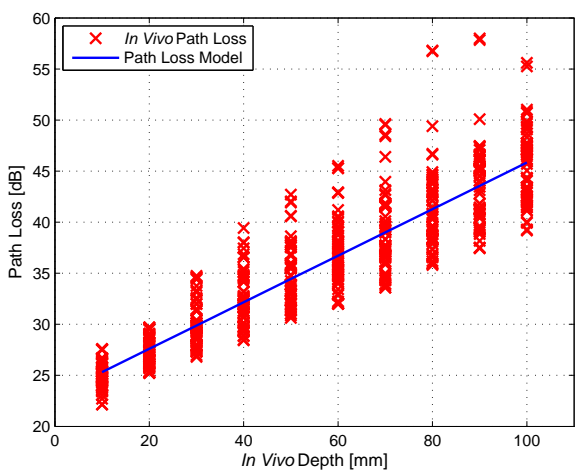}\caption{Scatter plot of path loss versus \textit{in vivo} depth at 915 MHz
\cite{demir}.\label{fig:scatter}}
\end{figure}

The channel modeling subgroup (Task Group 15.6), which worked on developing
of IEEE 802.15.6 standard, submitted their final report on body area
network (BAN) channel models in November 2010. In this report, it
is determined that Friis transmission equation can be used for \textit{in
vivo} scenarios by adding a random variation term and the path loss
was modeled statistically with a log-normal distributed random shadowing
$S$ and path loss exponent $n$ \cite{chavez2013propagation}. The
path loss exponent ($n$) heavily depends on environment and is obtained
by performing extensive simulations and measurements. In addition,
the shadowing term ($S$) depends on the different body materials
(e.g. bone, muscle, fat, etc.) and the antenna gain in different directions
\cite{Sayrafian}. The research efforts on assessing the statistical
properties of the \emph{in vivo} propagation channel are not finalized,
and there are still many open research efforts dedicated to building
analytical models for different body parts and operational frequencies
\cite{Alomainy,Sani,Sayrafian,Stoa,demir}.

\begin{table}[t]
\caption{Further details of the studies presented in Fig. 5.\label{tab:OpFreqs} }

\centering\includegraphics[width=0.9\columnwidth]{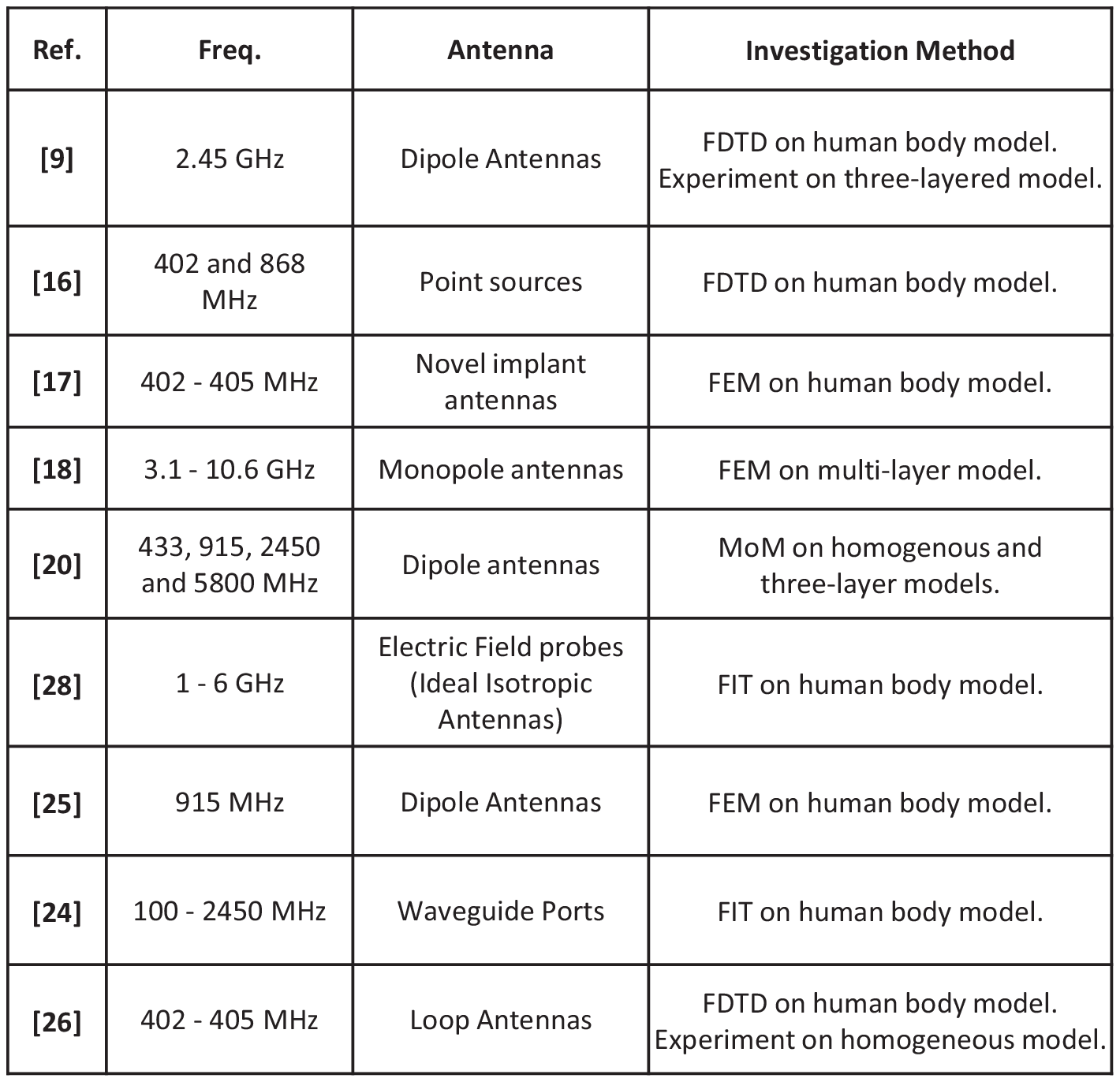}
\end{table}

A recent work investigates the \textit{in vivo} channel for the human
male torso at 915 MHz \cite{demir}. Fig. \ref{fig:scatter} shows
the scatter plot of path loss versus \textit{in vivo} depth in the
simulation environment. The \textit{in vivo} antenna is placed at
various locations (stomach area, intestine area, etc.) and various
depths (10 mm to 100 mm) inside the body and the \textit{ex vivo}
antenna is placed 5 cm away from the body surface. The path loss is
modeled as a function of depth by a linear equation in dB. The shadowing
presents a normal distribution for a fixed distance and its variance
becomes larger due to the increase in number of scattering objects
as the \textit{in vivo} antenna is placed deeper. The location-specific
statistical \textit{in vivo} path loss model parameters and a power
delay profile are provided in this study. The results confirm that
the \emph{in vivo} channel exhibits different characteristics than
the classical communication channels and location dependency is very
critical for link budget calculations.

\section{Open Research}

\textit{In vivo}-WBAN devices are expected to provide substantial
flexibility and improvement in remote healthcare by managing more
diseases and disabilities and their usage will likely increase in
time. Therefore, \textit{in vivo} channel characterization for a huge
variety of body parts is an obvious requirement for their future deployment
scenarios. With such models, wireless communication techniques can
be optimized for this environment and efficiently implemented. However,
research into solutions to satisfy emerging requirements for \textit{in
vivo}-WBAN devices such as high data rates, power efficiency, low
complexity, and safety should continue and continuous improvement
of channel characterization is necessary to optimize performance.

\begin{table*}[t]
\caption{A review of selected studied path loss models for various scenarios.
\label{tab:PL_models}}

\centering\includegraphics[width=1.6\columnwidth]{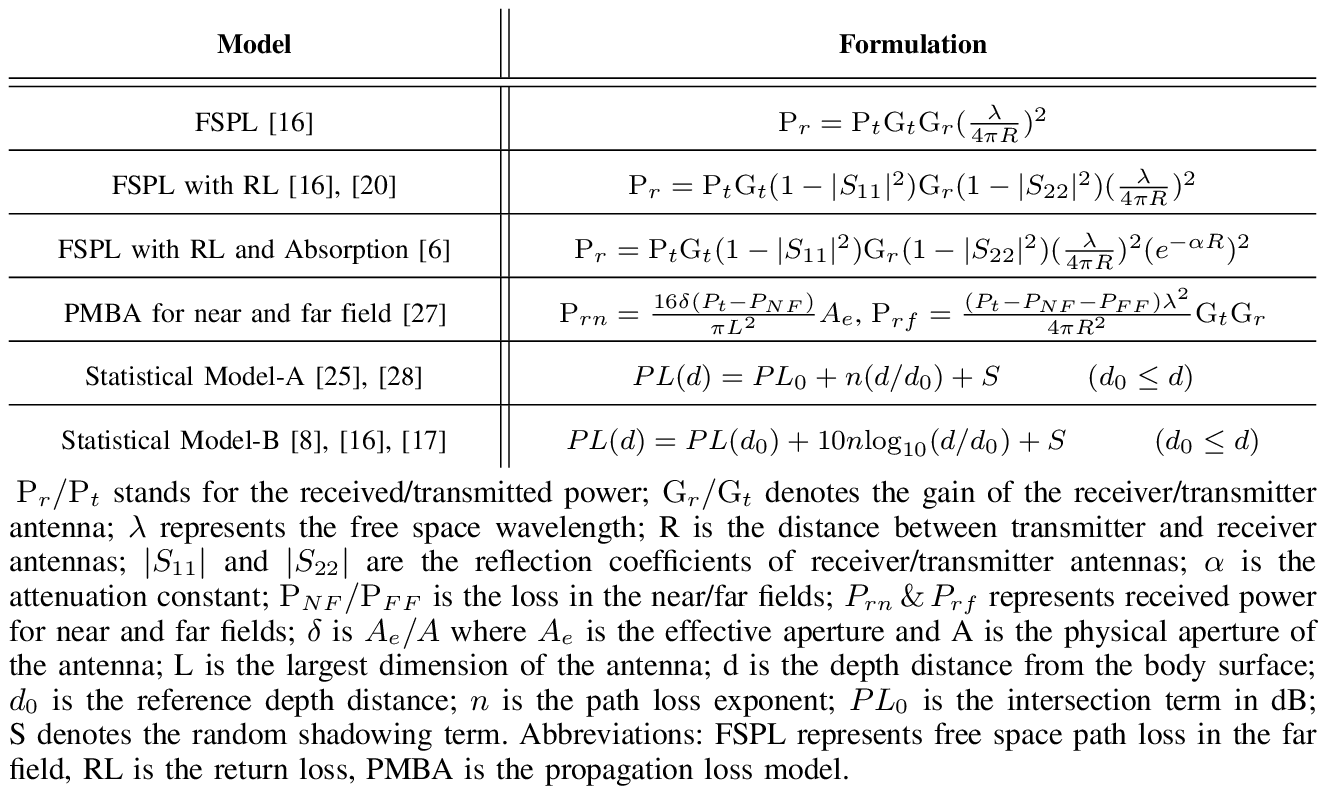}
\end{table*}

Some of the most important open research topics for efficient \textit{in
vivo} wireless communications are given as follows:
\begin{itemize}
\item \textit{Subject-Specific Studies}: It is known that on-body communication
channel is subject-specific \cite{Hall}. Additional studies need
to be performed on the subject-specific nature of \emph{in vivo} channels
to better understand the communication channel variations with respect
to the change of subject. This will help in developing efficient and
reliable implantable systems in future.
\item \textit{Security}: It is one of the most critical issues in the usage
of \textit{in vivo}-WBAN devices as various malicious attacks may
result in serious health risks, even death. Therefore, robust security
algorithms are essential for confidently using these devices. Physical
layer (PHY) security is a promising concept for providing security
in wireless communication \cite{ankaraliphysical}. Since most of
the proposed techniques in this field utilize the mutual channel information
between the legitimate transmitter and receiver, \textit{in vivo}
channel characterization considering the requirements of PHY-based
security methods is very important for implementing such techniques
on \textit{in vivo}-WBAN devices.
\item \textit{Multiple Input, Multiple Output (MIMO) and Diversity}: To
overcome ever increasing data rate demand and fidelity issues, while
keeping compactness in consideration for \emph{in vivo} communication,
MIMO and diversity based methods are very promising \cite{hevivo}.
However, the knowledge of spatial correlation inside the body medium
should be investigated for facilitating the implementation of these
techniques and understanding the maximum achievable channel capacity. 
\item \textit{Adaptive Communications}: Although, the \emph{in vivo} medium
is not as random as an outdoor channel, natural body motions and physiological
variations may lead to some changes in the channel state. Therefore,
more specific channel parameters, e.g., coherence time, coherence
bandwidth, Doppler spread in vivo medium should also be investigated
for facilitating adaptive communication against physical medium variations
to maintain adequate performance for specific scenarios under different
circumstances.
\item \textit{Interference and co-existence of WBAN devices }: Inter-WBAN
interference emerges as another problem for patients having multiple
\emph{in vivo}-WBAN sensors and actuators. Energy efficient techniques
enabling multiple closely located WBAN devices to co-exist are also
crucial for future applications and should be considered as an open
research.
\item \textit{Nanoscale in vivo wireless communication}: With the increase
in demand for compact and efficient implantable devices, nano-communication
technologies provide an attractive solution for potential BANs. More
studies are needed to better understand the \emph{in vivo} propagation
at terahertz frequencies, which is regarded as the most promising
future band for the electromagnetic paradigm of nano-communications.
In addition, studies are also needed to explore the connection between
micro-devices and nano-devices, which will be helpful for the design
of future system-level models.
\end{itemize}

\section{Conclusions}

In this paper, the state of the art of \textit{in vivo} wireless channel
characterization is presented. Various studies have been highlighted
from the literature for \textit{in vivo} channel models, considering
different parameters and by taking into account various anatomical
regions. A complete model is not available and remains as an open
research objective. However, considering the expected future growth
of implanted technologies and their potential use for the detection
and diagnosis of various health related issues in the body, the channel
modeling studies should be extended to enable the development of more
efficient communications systems for future \textit{in vivo} systems.

\bibliographystyle{IEEEtran}
\bibliography{inVivoMagazine}
\begin{IEEEbiography}[\includegraphics{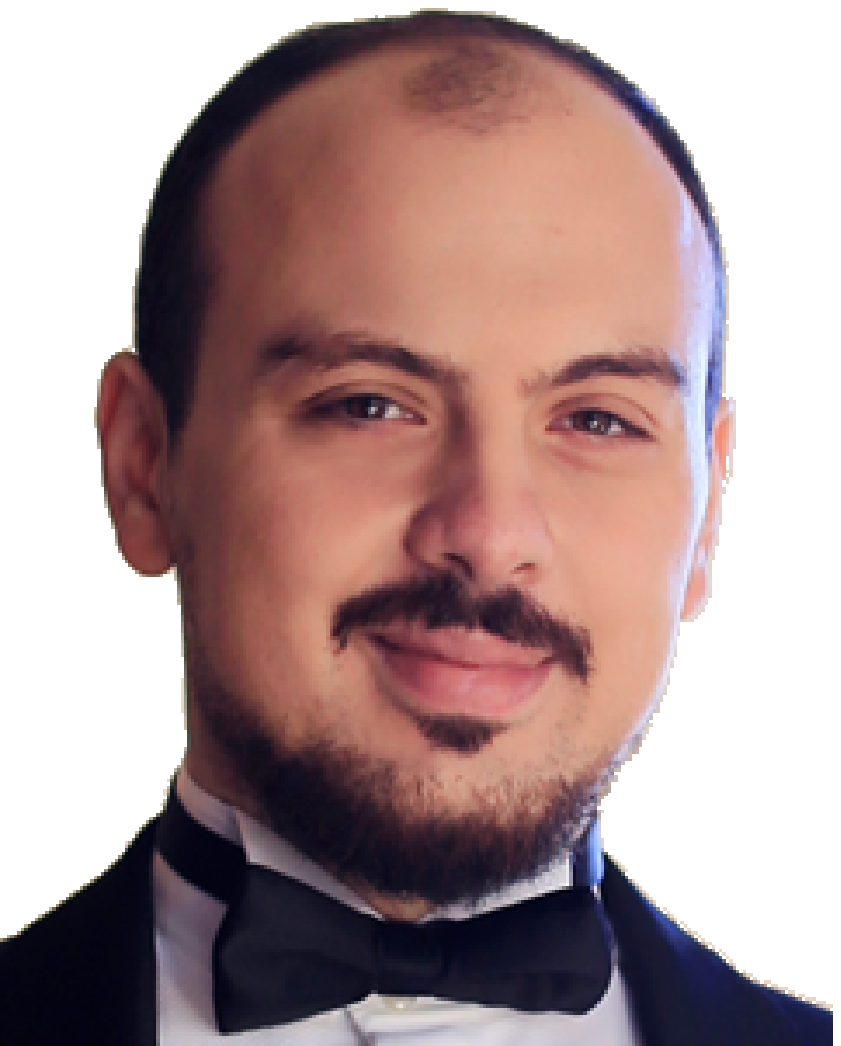}]{Ali Fatih Demir}
(S'08) received his B.S. degree in electrical engineering from Yildiz
Technical University, Istanbul, Turkey, in 2011 and his M.S. degrees
in electrical engineering and applied statistics from Syracuse University,
Syracuse, NY, USA in 2013. He is currently pursuing his Ph.D. degree
as a member of the Wireless Communication and Signal Processing (WCSP)
Group in the Department of Electrical Engineering, University of South
Florida, Tampa, FL, USA. His current research interests include \emph{in
vivo} wireless communications, biomedical signal processing, and brain-computer
interfaces. He is a student member of the IEEE.
\end{IEEEbiography}

\begin{IEEEbiography}[\includegraphics{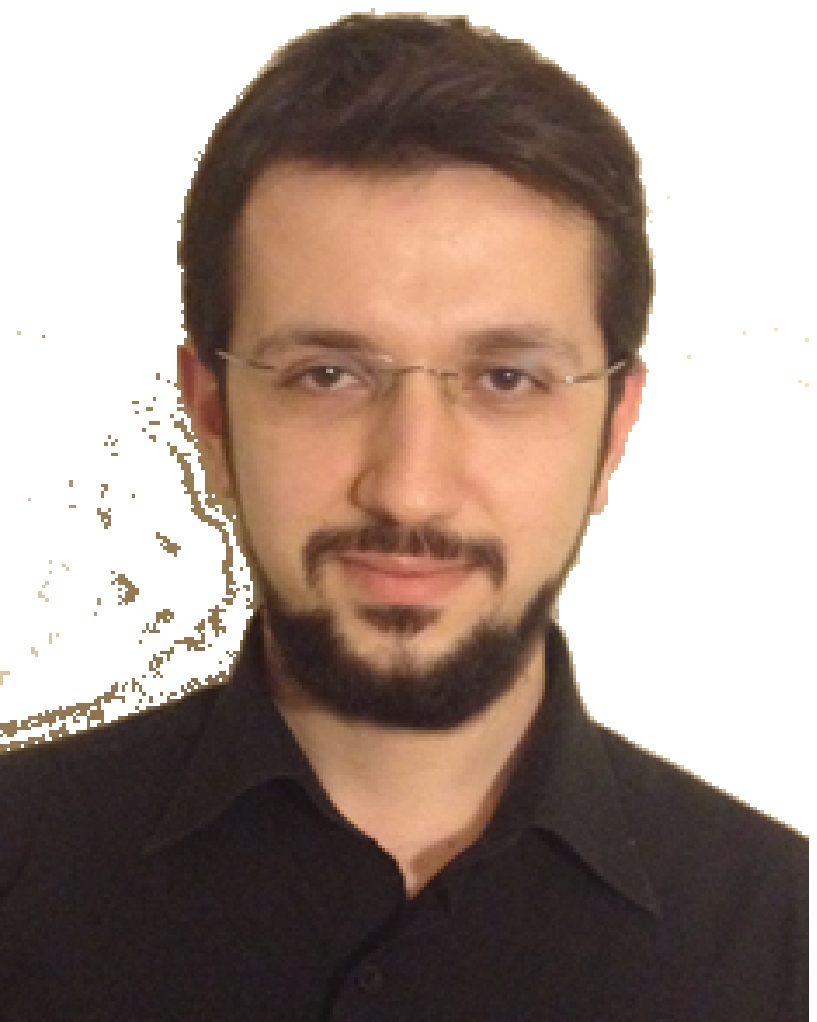}]{Z. Esat Ankarali}
(S'15) received his B.S. degree in control engineering from Istanbul
Technical University, Istanbul, Turkey, in 2011 and his M.S. degree
in electrical engineering from the University of South Florida, Tampa,
FL, USA, in 2013, where he is currently pursuing a Ph.D. degree as
a member of the WCSP Group in the Department of Electrical Engineering.
His current research interests include multi-carrier systems, physical
layer security, and \emph{in vivo} wireless communications. He is
a student member of the IEEE.
\end{IEEEbiography}

\begin{IEEEbiography}[\includegraphics{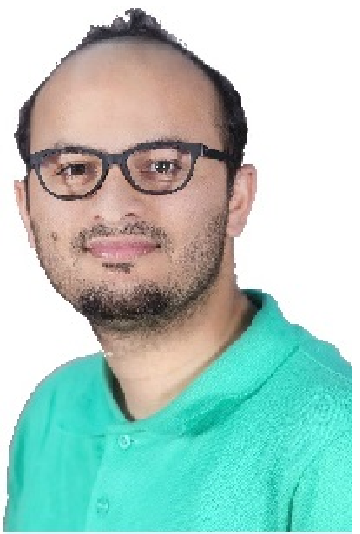}]{Qammer H. Abbasi}
 (S'08\textendash M'12\textendash SM'16) received his B.S. degree
in electronics engineering from the University of Engineering and
Technology, Lahore, Pakistan in 2007, his Ph.D. degree in electronic
and electrical engineering from Queen Mary University of London, U.K.,
in 2012 where he has been a visiting research Fellow since 2013. He
joined the Department of Electrical and Computer Engineering (ECEN)
of Texas A\&M University at Qatar in August 2013, where he is now
an assistant research scientist. His research interests include compact
antenna design, radio propagation, body-centric wireless communications,
cognitive/cooperative network, and MIMO systems. He is a member of
IET and a senior member of the IEEE.
\end{IEEEbiography}

\begin{IEEEbiography}[\includegraphics{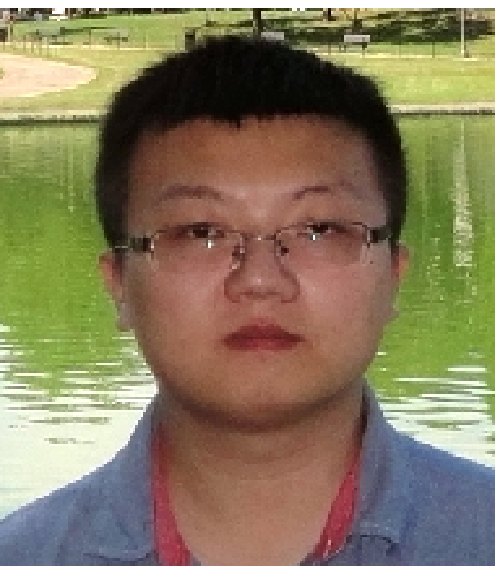}]{Yang Liu}
  received the B.S. degree in Biology Science from Wuhan University
in 2010 and the M.S. degree in Electrical Engineering from Beijing
University of Posts and Telecommunications in 2013. Currently, he
is pursuing the Ph.D. degree in Electrical Engineering in University
of South Florida. His research interests include in vivo wireless
communications and networking.. 
\end{IEEEbiography}

\begin{IEEEbiography}[\includegraphics{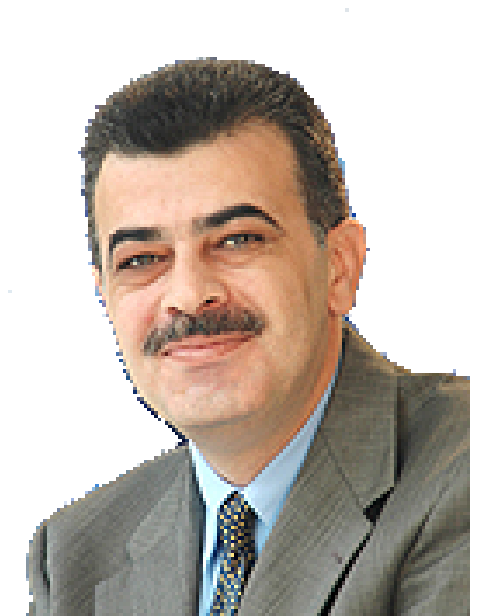}]{Khalid Qaraqe}
 (M\textquoteright 97\textendash SM\textquoteright 00) received
his B.S. degree in electrical engineering (EE) from the University
of Technology, Baghdad, Iraq, in 1986. He received his M.S. degree
in EE from the University of Jordan, Amman, in 1989, and he received
his Ph.D. degree in EE from Texas A\&M University, College Station,
TX, USA, in 1997. He joined the Department of ECEN of Texas A\&M University
at Qatar in July 2004, where he is now a professor. His research interests
include communication theory, mobile networks, cognitive radio, diversity
techniques, and beyond fourth-generation systems. He is a senior member
of the IEEE.
\end{IEEEbiography}

\begin{IEEEbiography}[\includegraphics{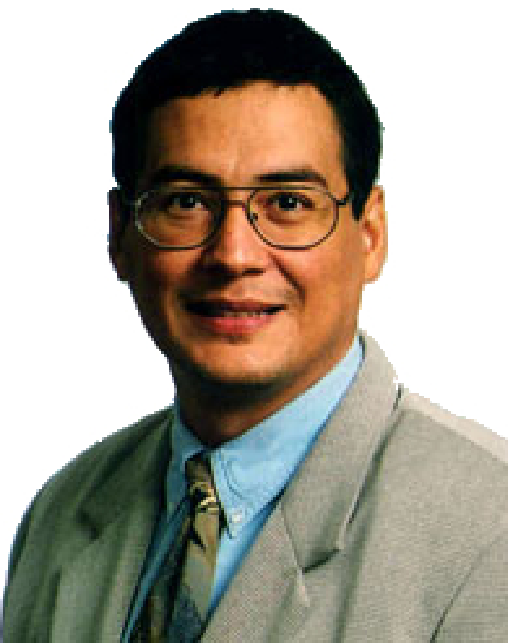}]{Erchin Serpedin}
 (S\textquoteright 96\textendash M\textquoteright 99\textendash SM\textquoteright 04\textendash F\textquoteright 13)
received his specialization degree in signal processing and transmission
of information from Ecole Superieure D\textquoteright Electricite,
Paris, France, in 1992, his M.S. degree from the Georgia Institute
of Technology, Atlanta, GA, USA in 1992, and his Ph.D. degree in electrical
engineering from the University of Virginia, Charlottesville, VA,
USA, in January 1999. He is a professor with the Department of ECEN,
Texas A\&M University, College Station, TX, USA. He is serving as
an associate editor of IEEE Signal Processing Magazine and as the
editor-in-chief of European Association for Signal Processing Journal
on Bioinformatics and Systems Biology. His research interests include
signal processing, biomedical engineering, and machine learning. He
is a fellow of the IEEE. 
\end{IEEEbiography}

\begin{IEEEbiography}[\includegraphics{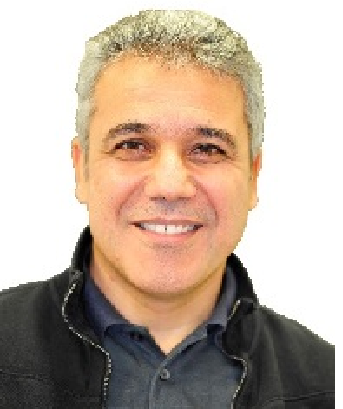}]{Huseyin Arslan}
 (S\textquoteright 95\textendash M\textquoteright 98\textendash SM\textquoteright 04\textendash F\textquoteright 16)
received his B.S. degree from Middle East Technical University, Ankara,
Turkey, in 1992 and his M.S. and PhD. degrees from Southern Methodist
University, Dallas, TX, USA, in 1994 and 1998, respectively. From
January 1998 to August 2002, he was with the research group of Ericsson
Inc., NC, USA, where he was involved with several projects related
to 2G and 3G wireless communication systems. Since August 2002, he
has been with the Department of Electrical Engineering, University
of South Florida, Tampa, FL, USA, where he is a Professor. In December
2013, he joined joined Istanbul Medipol University, Istanbul, Turkey,
where he has worked as the Dean of the School of Engineering and Natural
Sciences. His current research interests include waveform design for
5G and beyond, physical layer security, dynamic spectrum access, cognitive
radio, coexistence issues on heterogeneous networks, aeronautical
(high altitude platform) communications, and \emph{in vivo} channel
modeling and system design. He is currently a member of the editorial
board for the Sensors Journal and the IEEE Surveys and Tutorials.
He is a fellow of the IEEE.
\end{IEEEbiography}

\end{document}